\documentclass[floatfix, prb, twocolumn, showpacs]{revtex4}

\usepackage{graphicx}

\usepackage{dcolumn}

\usepackage{bm}

\begin{document}

\preprint{\today}
\title{Magneto-optic far-infrared study of Sr$_{14}$Cu$_{24}$O$_{41}$: triplet excitations in chains. 
}
\author{D.~H{\"u}vonen}
\email{dan@kbfi.ee}
\author{U.~Nagel}
\author{T.~R{\~o}{\~o}m}
\affiliation{National
Institute of Chemical Physics and Biophysics,
    Akadeemia tee 23, 12618 Tallinn, Estonia.}

\author{P.~Haas}
\author{M.~Dressel}
\affiliation{1. Physikalisches Institut, Universit{\"a}t Stuttgart, D-70550 Stuttgart, Germany}

\author{Y.-J.~Wang}
\affiliation{
National High Magnetic Field Laboratory, Florida State University, 
1800 East Paul Dirac Drive, Tallahassee, Florida 32306, USA}

\author{J.~Akimitsu}
\affiliation{Department of Physics, Aoyama-Gakuin University, Tokyo, Japan}


\begin{abstract}
Using far-infrared spectroscopy we have studied 
the magnetic field and temperature dependence 
of the spin gap modes in the chains of Sr$_{14}$Cu$_{24}$O$_{41}$.
Two triplet modes T$_1$ and T$_2$ were found in the center of the Brillouin zone
at $\Delta_1=9.65$\,meV and $\Delta_2=10.86$\,meV in zero magnetic field.
The T$_1$ mode was excited when the electric field vector ${\bf E}$ of the light
was polarized along the $b$ axis (perpendicular to the planes of chains and ladders) and
T$_2$ was excited for ${\bf E}\parallel {\bf a}$ (perpendicular to the chains
and along the rungs).
Up to the maximum magnetic field of 18\,T,  
applied along the chains, the electron $g$ factors of 
these two modes were similar,
$g_{1c}=2.049$ and $g_{2c}=2.044$.
Full linewidth at half maximum for both modes 
was 1\,cm$^{-1}$ (0.12\,meV) at 4K and increased with $T$.
The temperature dependence of mode energies and line intensities was in agreement with the inelastic neutron scattering results from two groups [Matsuda {\it et al.}, Phys. Rev. B {\bf 59}, 1060 (1999) and Regnault {\it et al.}, Phys. Rev. B {\bf 59}, 1055 (1999)]. 
The T$_1$ mode has not been observed by inelastic neutron scattering  in the points of the $k$-space 
equivalent to the center of the Brillouin zone.
Our study indicates that the zone structure model of magnetic excitations of Sr$_{14}$Cu$_{24}$O$_{41}$ must be modified
 to include a triplet mode at 9.65\,meV in the center of the magnetic Brillouin zone.

\end{abstract}


\maketitle

\section{Introduction}

Heisenberg spin 1/2 systems have been investigated extensively both by experimental and theoretical means due to their versatile low-energy physical properties 
and also because of their relevance to high-$T_{c}$ superconducting materials.
The search for high-$T_{c}$ superconductor materials has led to a new structure type\cite{mccarron88,siegrist88} represented by Sr$_{14}$Cu$_{24}$O$_{41}$ 
containing both one-dimensional CuO$_{2}$ spin chains and two-dimensional Cu$_{2}$O$_{3}$ spin ladders.
Planar chains and ladders in this compound extend in $c$ axis direction and are alternately stacked along the $b$ axis, separated by layers of Sr. 
Chain and ladder spin subsystems in Sr$_{14}$Cu$_{24}$O$_{41}$ interact weakly and are structurally incommensurate although the lattice constants in $c$ direction satisfy an approximate relation $10\,c_{\rm chain} \approx 7\,c_{\rm ladder}$.
The conductivity in Sr$_{14}$Cu$_{24}$O$_{41}$ is associated with the charge dynamics in the ladder layers.
Sr substitution for Ca and external pressure leads to hole transfer from the chains to the ladders and to the occurrence of superconductivity in Sr$_{0.4}$Ca$_{13.6}$Cu$_{24}$O$_{41}$
 with $T_c= 12$\,K at 3\,GPa.\cite{uehara96}
For a comprehensive overview on the charge and spin dynamics in this class of materials the reader is directed to Ref. \onlinecite{vuletic06}.
Pure Sr$_{14}$Cu$_{24}$O$_{41}$ is a self-doped compound containing six holes per unit cell. 
A chain hole occupies oxygen $2p$ orbitals surrounding a central Cu spin and forms a Zhang-Rice (ZR) singlet\cite{zhang88} (-0-), rendering nonmagnetic about 6 out of 10 Cu sites.
Spin dimers, two Cu$^{2+}$ spins bridged by a ZR singlet (-$\uparrow$-0-$\downarrow$-),\cite{matsuda96,carter96,takigawa98} are in the singlet state. 
Inelastic neutron scattering,\cite{matsuda99,regnault99,eccleston98} high energy X-ray diffraction\cite{fukuda02} and NMR\cite{takigawa98} measurements have indicated that dimers organize with a periodicity of 5 chain units and are separated by two ZR singlets (-$\uparrow$-0-$\downarrow$-0-0-).

Recent revised structural studies indicate extensive O atom position modulation out of the chain planes in  Sr$_{14}$Cu$_{24}$O$_{41}$\cite{smaalen03, braden04, etrillard04, gotoh03, gotoh06} and  Sr$_{0.4}$Ca$_{13.6}$Cu$_{24}$O$_{41}$.\cite{isobe00}
This modulation causes variations in the super-exchange between Cu atoms along the chain.\cite{gelle04,gelle05}
In addition, displaced O atoms mediate the hole transfer between chains and ladders.
It is estimated from bond-valence sum calculations, X-ray absorption spectroscopy, magnetization and optical conductivity measurements\cite{gotoh03,nucker00,carter96,osafune97,kataev01,klingeler06} that 1-4$\%$ of the self-doped holes reside in the ladders in undoped Sr$_{14}$Cu$_{24}$O$_{41}$ at low temperature. 
The deficiency of holes in the chains means that the perfect alignment of dimers, separated by two ZR singlets yielding five chain unit periodicity, cannot be satisfied.
There is a lack of consensus in the literature regarding spin and charge order in the chains of Sr$_{14}$Cu$_{24}$O$_{41}$ and the nature of the underlying ground state.
We studied magnetic excitations using far-infrared (FIR) spectroscopy and strong magnetic fields with the aim to identify the spin states present in the chains of Sr$_{14}$Cu$_{24}$O$_{41}$.

\section{Experimental}

FIR transmission spectra were recorded with a polarizing Martin-Puplett type Fourier transform spectrometer SPS200. 
Samples,  a 12\,T superconducting magnet, and two 0.3\,K silicon bolometers were inside a $^{4}$He cryostat connected to the spectrometer through  light pipe.
A rotatable polarizer was placed in front of the sample.
Absorption spectra were calculated considering two back reflections from the sample.
Spectra in fields above 12\,T were measured at NHMFL in Tallahassee utilizing Bruker IFS\,113v with a 18\,T superconducting magnet and a 4\,K silicon bolometer.
Two single-crystalline samples of Sr$_{14}$Cu$_{24}$O$_{41}$ were used in the current study: a 1.1\,mm thick crystal with an $(ab)$-plane area of 13.2\,mm$^{2}$ and an $(ac)$-plane crystal with an area of 12.6\,mm$^{2}$ and a thickness of 0.65\,mm. 

\section{Results}

\begin{figure}[tb]
\includegraphics[width=8.6cm]{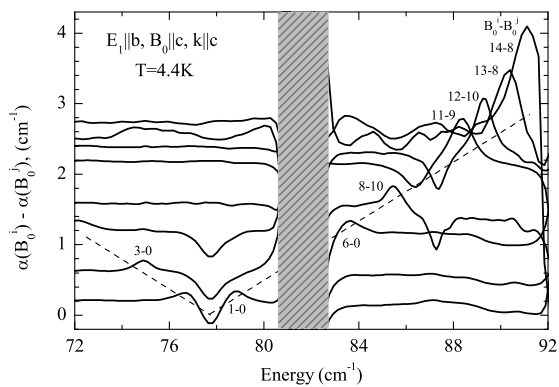}
\caption{Differential absorption spectra of  triplet T$_{1}$ in magnetic field applied in the $c$ axis direction with light $k$ vector $\mathbf{k}\parallel\mathbf{c}$
  and polarization $\mathbf{E}_{1}\parallel\mathbf{b}$ at 4.4\,K. 
The spectra are shifted in the vertical direction by $0.2B_0^i$.
Dashed lines are eye guides.
The shaded area covers the spectral region of strong phonon absorption.}
\label{specEpB}
\end{figure}

\begin{figure}[tb]
\includegraphics[width=8.6cm]{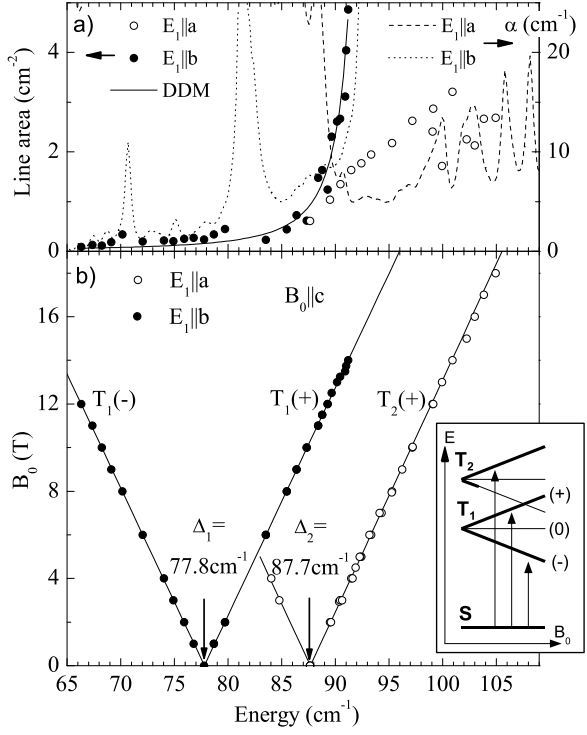}
\caption{ Magnetic field  $\mathbf{B_0}\parallel  \mathbf{c}$ dependence of line areas and positions.
 a) Line areas of triplet modes (left axis) and FIR absorption spectra (right axis) at temperature 4.4\,K for  $\mathbf{E}_{1}  \parallel \mathbf{a}$ and $\mathbf{E}_{1}  \parallel  \mathbf{b}$ polarizations.
The solid line is the dynamic Dzyaloshinskii-Moriya interaction model fit of the T$_1$ absorption line area. 
b)  T$_1$ and T$_2$ line positions as a function of magnetic field $B_{0}$. 
The inset shows schematically the observed transitions.}
\label{bothtrips}
\end{figure}

Polarization-sensitive transmission measurements in the FIR spectral region revealed an anisotropic response from the crystal $(ab)$ plane. 
When the external magnetic field was applied along the $c$ axis of the crystal (along the chains),
two magnetic field dependent modes, T$_{1}$ and T$_{2}$ were found, which we assign to spin excitations in the chains of Sr$_{14}$Cu$_{24}$O$_{41}$. 
Transitions to the triplet state T$_{1}$ were visible when the electric field component $\mathbf{E}_{1}$ of the radiatioan was polarized along the $b$ axis. 
Observed differential absorption spectra, measured in different magnetic fields at 4\,K, are shown in Fig.\,\ref{specEpB}. 
In the spectra reference field lines are pointing downward and lines in the measured fields are pointing upward.  
The triplet state has three spin sublevels, $M_S=-1,0,+1$ what we denote as  T$(-)$, T$(0)$, and  T$(+)$, see inset to Fig.\,\ref{bothtrips}.
The resonance frequency of the T$_{1}(0)$ level does not shift with field and thus escapes detection.
The transition to the T$_{1}(-)$ sublevel loses intensity as the line shifts toward smaller energies in increasing field and gets too weak for detection in fields above 12\,T. 
The transition to the T$_{1}(+)$ level gains intensity with increasing magnetic field up to the observation limit set by the strong phonon background at 92 cm$^{-1}$ where the crystal becomes opaque to FIR radiation in $\mathbf{E}_{1}\parallel \mathbf{b}$ polarization. 

In $\mathbf{E}_{1}\parallel \mathbf{a}$ polarization transitions to the triplet state T$_{2}$ are observed (Fig.\,\ref{specEpA}). 
Transitions to the T$_{2}(+)$ level in fields below 2\,T and to the T$_{2}(-)$ level are masked by strong absorption below 89\,cm$^{-1}$.
To check the light polarization and $\mathbf{B}_{0}$ orientation dependence, measurements  on a  thinner $(ac)$-plane crystal in Faraday and  Voigt configuration were carried out. 
When $\mathbf{B}_{0}\parallel \mathbf{b}$  the triplet T$_{2}$ mode disappears from the infrared absorption spectrum.
Some absorption lines belonging to the T$_{2}(-)$ level were detected  in Voigt configuration. 
However, because of strong phononic absorption the intensities of T$_{2}(-)$ were not reliable and are not included in Fig\,\ref{bothtrips}a.

The spectral lines were fitted with a Lorentzian line shape. Line positions and intensities as a function of magnetic field are displayed in Fig.\,\ref{bothtrips}. 
We assume that T$_{1}(-)$ and T$_{1}(+)$ are degenerate at $B_0=0$\,T and plot half of the measured line intensity for this field value. 
The same holds for the T$_{2}$  triplet.
Linewidths of both triplets are $1\pm0.2$ cm$^{-1}$ and do not depend on the strength of the magnetic field.
The corresponding $g$ factors are similar, $g_{1c}=2.049\pm 0.012$ and $g_{2c}=2.044\pm  0.014$.
Additionally we observed a strong paramagnetic signal at 4\,K with $g_{c}=2.038\pm0.016$. 
The linewidth of the paramagnetic signal was below the used instrumental resolution limit, 0.3\,cm$^{-1}$. 

\begin{figure}[tb]
\includegraphics[width=8.6cm]{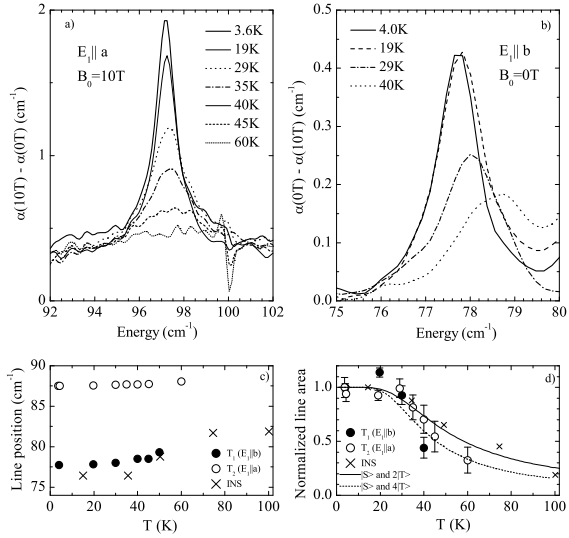}
\caption{Temperature dependence of triplet modes: a)  T$_{2}(+)$ in 10\,T field, b) T$_{1}$ in zero magnetic field, c) line positions, 
d) observed normalized transition intensities;  calculated singlet-triplet transition line area assuming 2 triplets  (solid line) or 4 triplets in the folded zone structure (dashed line). 
The INS data, Ref. \onlinecite{matsuda99}, is at $\mathbf{k}=(2,0,-0.1)$.}
\label{tdep}
\end{figure}

The temperature dependence of the singlet-triplet resonances was measured  as the difference of  0\,T and 10\,T spectra at temperatures from 4\,K to 60\,K, see Fig.\,\ref{tdep}.
Fig.\,\ref{tdep}b shows absorption lines in 0\,T field  for triplet T$_1$.
Sr$_{14}$Cu$_{24}$O$_{41}$ is not transparent in $\mathbf{E}_{1}\parallel \mathbf{a}$ polarization at zero field line position frequencies of T$_2$ and therefore we analyze the $T$ dependence of this triplet level in 10\,T field. 
Although both lines broaden and lose intensity as temperature increases, the  $T$ dependence of line positions is different. 
The T$_{1}$ triplet resonance shifts notably toward higher energies whereas the energy of T$_{2}$ remains unchanged, see Fig.\,\ref{tdep}c.

\section{Discussion}

\begin{figure}[dispersions]
\includegraphics[width=8.6cm]{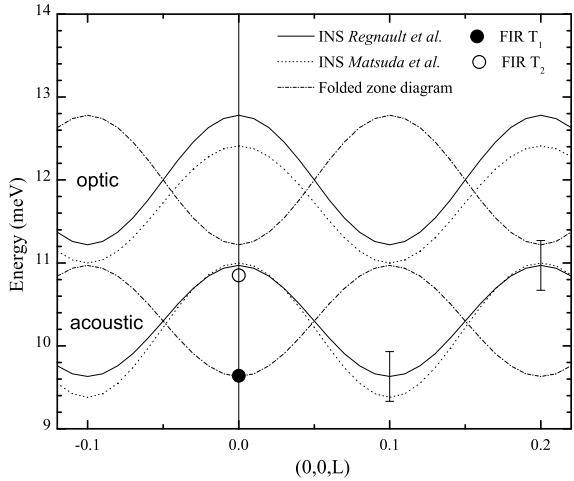}
\caption{Dispersion curves of the magnetic excitations in the chains (Refs. \onlinecite{matsuda99} and \onlinecite{regnault99}) and the observed triplet modes from FIR.
The dash-dot lines are extra dispersion curves, derived from data by {\sl Regnault et al.}\cite{regnault99}, 
after doubling the unit cell in the $c$ axis direction.
INS energy resolution is shown by error bars; FIR transitions have linewidth less  than the size of experimental points on the graph.}
\label{disp}
\end{figure}

The dispersion curves of two magnetic excitations, acoustic and optic, with respective energies 11\,meV and 12.5\,meV at $k$ points equivalent to $\mathbf{k}=0$ have been measured by inelastic neutron scattering (INS) spectroscopy.\cite{matsuda99, regnault99}
A good fit of the experimental data was obtained by the simplest model for a weakly coupled dimer system.
The dispersion in the $c$ axis direction follows a cosine form with the periodicity of 0.2 reciprocal lattice units (Fig.\,\ref{disp}). 
The temperature dependence of INS data demonstrated that with rising temperature the dispersion curves flatten out since the inter-dimer couplings become negligible due to thermal fluctuations and the dimers behave more like isolated dimers. 
As the dispersion curves flatten, the acoustic mode at $k$ space points near the Brillouin zone edge $(H,K,0.1)$ shift toward higher energies while the triplet state energy at $(H,K,0)$ does not change (reciprocal lattice units $H$ and $K $ are integers in our discussion).
In contrast to these considerations, the energy of the optic branch at $(H,K,0.1)$ in the momentum space displays no temperature dependence while the energy near $(H,K,0)$ lowers notably with temperature.

It follows from the momentum conservation that an absorbed infrared photon creates a triplet excitation with $\mathbf{k}$ equal to photons momentum,  $\mathbf{k}\approx 0$,
this is in the center of Brillouin zone.
INS data\cite{matsuda99, regnault99} shows no excitation at 9.65\,meV in the center of the Brillouin zone that would correspond to the triplet excitation T$_{1}$, seen in FIR spectra.
A possible explanation to overcome this discrepancy with INS results is to consider zone folding.
Doubling of the unit cell size in the $c$ axis direction is consistent with FIR data since then the energy of the back folded acoustic branch at $(H,K,0.1)$ point would coincide with the energy of  T$_{1}$, see Fig.\ref{disp}.
Indeed, the triplet mode T$_{1}$ has a similar temperature dependence of its energy as the acoustic mode at $k$ points equivalent (any integer value of $H$ and $K$)  to $(H,K,0.1)$.
We conclude that a magnetic unit cell of 10\,$c_{\rm chain}$ is compatible with our results.

The triplet T$_{2}$ has an energy similar to the acoustic mode at (H,K,0).
In the folded zone picture the (H,K,0.1) point moves to the center of the Brillouin zone and is infrared-active as well.
Due to the experimental uncertainty of the INS experiment ($\sim$\,0.5 meV) and a similar temperature dependence of acoustic and optic modes at this energy ($\approx 11$\,meV) we cannot determine exactly whether the triplet T$_{2}$ belongs to the acoustic or the zone folded optic branch of excitations, seen by INS.
However, the $k$ points equivalent to $(H,K,0)$ points of the acoustic excitation are closer in the energy scale to the triplet level T$_{2}$, seen by FIR.
Consequently we assign triplet T$_{2}$ to the chain acoustic excitation branch (in INS notation) at $\mathbf{k}=0$ point in the momentum space and refine the spin gap value for this point to 10.86\,meV (87.7\,cm$^{-1}$). 

The $T$ dependence of intensity of the observed FIR transitions, $I$,  is proportional to the population difference of singlet and triplet levels if we assume that the transition matrix element is independent of temperature.
Following the Boltzmann distribution we get $I \sim [1-\exp(-\Delta/kT)]/[1+3N \exp(-\Delta/k_{B}T)]$, where $N$ is the number of triplet states, each 3-fold degenerate, and $\Delta$ is the singlet-triplet energy gap; we used an averaged value $\Delta=11$\,meV  for all triplets.
The solid line (Fig.\,\ref{tdep}d) indicates the calculated  normalized transition intensity for a system with two  triplet states and the dashed line for four triplet states in the new folded zone structure.
Theoretical curves  qualitatively explain the decrease in intensity.
As the lines broaden with temperature, the determination of line area becomes less accurate and thus $T$ dependence of intensity cannot distinguish whether two or four triplet states exist.

In the FIR we observed two distinct magnetic signals at 4\,K. 
Firstly, the transitions from the singlet ground state to triplet states T$_{1}$ and T$_{2}$, and secondly, a paramagnetic signal.
It is not expected that the paramagnetic response at 4\,K originates 
from a thermally excited triplet state in the chains due to a large spin gap, $\Delta\approx 11$\,meV. 
There is evidence from susceptibility and ESR measurements that unpaired spins exist in Sr$_{14}$Cu$_{24}$O$_{41}$.
The $g$ factor of the paramagnetic signal $g_{c}=2.038\pm0.016$ obtained from our experiments matches that of ESR $g_{c}(4\mathrm{K})$=2.038 at 4\,K.\cite{klingeler06}
Above 20\,K when the triplets are thermally populated, the
ESR detects transitions between triplet levels $M_S=\pm 1$ and $M_S=0$.
The $g$ factors of both triplets measured in FIR, $g_{1c}=2.049$ 
and $g_{2c}=2.044$,
coincide with averaged value $g_{c}(20\mathrm{K})=2.045$ from ESR.\cite{klingeler06,matsuda96,kataev01}
FIR measurements confirm directly  that unpaired spins have $g_c$ similar to spin dimers  $g$ factors $g_{1c}$ and $g_{2c}$.
Similar $g_c$ components of the $g$ tensor suggest a common local environment, as was proposed in Ref.\,\onlinecite{klingeler06},
for the triplet and the paramagnetic site magnetic moments.
Unpaired spins are in chains, instead of ladders.
Unpaired Cu$^{2+}$ spins in the chain subsystem destroy the perfect charge order with $5c_{\rm chain}$ periodicity and are in accord with the FIR active triplet T$_{1}$ which, as we discussed above, is a result of doubling of the unit cell. 

At 20\,K the linewidths of the singlet to triplet transitions seen in FIR is about 200 times larger than the ESR transitions,\cite{kataev01} $1\,\mathrm{cm}^{-1} / 200 = 5\times 10^{-3}\,\mathrm{cm}^{-1}\approx 50\,\mathrm{Oe} $.
The difference between the two experimental probes is that FIR transitions are across the spin gap while ESR transitions are within the triplet state.
It is natural to assume that it is the distribution of spin gaps what makes the 200-fold increase of triplet linewidth in FIR spectroscopy compared to ESR linewidth.
Indeed, theoretical calculations \cite{gelle05} have demonstrated that there is a substantial fluctuation, 8\%,  of intradimer exchange coupling, $\sim 10$\,meV, which contributes the most to the size of the spin gap.
This inhomogeneous distribution should give a Gaussian and not a Lorentzian lineshape.
However, there is inter-dimer exchange, $\sim 1$\,meV,  that reduces the effect of inhomogeneous broadening and restores the Lorentzian lineshape, an effect similar to motional narrowing.

Now we discuss selection rules of the observed singlet-triplet transitions. 
We have determined by measuring the crystals in different experimental geometries that both singlet-triplet transitions are electric dipole active. 
$\mathbf{E}_{1}\parallel \mathbf{b}$ polarized light excites the triplet state T$_{1}$ and $\mathbf{E}_{1}\parallel \mathbf{a}$ light promotes singlets to the T$_{2}$ triplet state. 
The electric field component of light cannot directly couple to a spin system. 
In order to explain our observations we consider a spin-phonon coupling mechanism where light couples to a phonon mode that lowers the symmetry of the lattice and generates an antisymmetric Dzyaloshinskii-Moriya (DM) interaction which couples singlet and triplet states.
This dynamic DM transition mechanism has been successful in describing observed electric dipole active transitions in SrCu$_{2}$(BO$_{3}$)$_2$ \cite{cepas04, room2004borate} and $\alpha$'-NaV$_{2}$O$_{5}$. \cite{room2004NaVa}

\begin{figure}[tb]
\includegraphics[width=8.6cm]{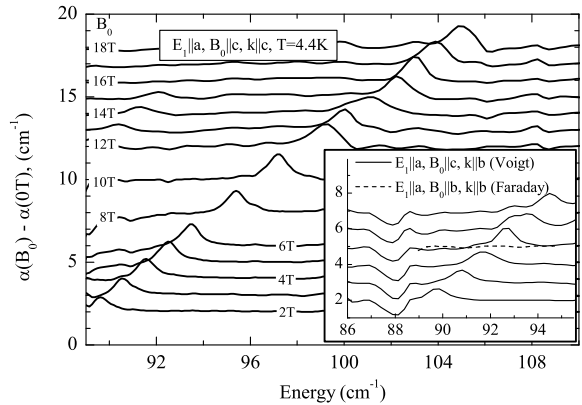}
\caption{Differential absorption spectra of  triplet T$_{2}$ in magnetic field applied in the $c$ axis direction with light $k$ vector $\mathbf{k}\parallel\mathbf{c}$
and polarization $\mathbf{E}_{1}\parallel \mathbf{a}$ at 4.4\,K measured using the ($ab$)-plane crystal. 
The spectra are offset in vertical direction by magnetic field value $B_0$. 
The inset shows differential absorption spectra for the ($ac$)-plane crystal in Faraday (dashed line) and Voigt (solid line) configurations. 
Triplet T$_2(+)$ is not infrared-active when $\mathbf{B}_{0}   \parallel  \mathbf{b}$
as shown by the dashed line for an arbitrary chosen $B_{0}=  5 $\,T.}
\label{specEpA}
\end{figure}

There are several observations that suggest a dynamic DM interaction in spin chains of Sr$_{14}$Cu$_{24}$O$_{41}$.
Firstly, the dependence on the light polarization.  The
dynamic DM vector $\mathbf{D}_{q}$ is perpendicular to $\mathbf{E}_{1}$ and to the spin dimer axis that is the $c$ axis in Sr$_{14}$Cu$_{24}$O$_{41}$.
The triplet T$_1$ interacts with electric field vector of the light $\mathbf{E}_{1}\parallel \mathbf{b}$ and $\mathbf{D}_{q}$  is parallel to $a$ axis for this triplet. 
The triplet T$_2$ interacts with the polarization $\mathbf{E}_{1}\parallel \mathbf{a}$ and $\mathbf{D}_{q}$  is parallel to $b$ axis. 
The selection rules are: i) if  $\mathbf{B}_{0}\parallel \mathbf{D}_{q}$ then the transition to T$(0)$ is allowed and 
ii) if $\mathbf{B}_{0}\perp \mathbf{D}_{q}$ then transitions to T$(-)$ and to T$(+)$ are allowed. 
One can see from Figs.\,\ref{specEpB} and \ref{specEpA} that the selection rules are satisfied:
 T$(-)$ and T$(+)$ transitions are observed when $\mathbf{B}_{0}\parallel \mathbf{c}$, selection rule ii) is satisfied.
Moreover, when the magnetic field was aligned along the $b$ axis then the T$_2(+)$ line disappeared in $\mathbf{E}_{1}\parallel \mathbf{a}$ geometry (Fig.\,\ref{specEpA}).
This behavior is consistent with the dynamic DM mechanism selection rule i).
Secondly, the dependence of transition intensities on magnetic field. 
According to Eq.\,\ref{Eq_intensity} a shift in energy of the triplet level reduces (increases) the separation between the triplet level and the relevant phonon mode  and increases (decreases) coupling between the triplet and phonon states. 
The magnetic field dependence of the intensity of the triplet T$_{1}$ excitation, Fig.\,\ref{bothtrips}a, was fitted with singlet-triplet transition probabilities for the dynamic DM model using Eq.\,13 and 14 from Ref. \onlinecite{room2004NaVa}:
\begin{eqnarray}
| \langle \mathrm{T}(-) | V | \mathrm{S} \rangle  |^2 = I_p \frac {(qD_q)^2 (\hbar \omega_{ph} )^2}
{2[(\hbar\omega_{ph})^2 - (\Delta -g\mu _B B_0)^2]^2}\,,\nonumber\\
| \langle \mathrm{T}(+) | V | \mathrm{S} \rangle  |^2 = I_p \frac {(qD_q)^2 (\hbar \omega_{ph} )^2}
{2[(\hbar\omega_{ph})^2 - (\Delta +g\mu _B B_0)^2]^2}\,.
\label{Eq_intensity}
\end{eqnarray}
The spin gap value $\Delta=\Delta_{1}=77.8$\,cm$^{-1}$ and the $g$ factor $g=g_c=2.049$ were fixed parameters; the phonon frequency, $\omega_{ph}$, and $I_p (qD_q)^2$ were the fitting parameters.
The fit converged to a phonon at $\omega_{ph}=94\pm0.3$\,cm$^{-1}$.
A phonon has been detected at 95 cm$^{-1}$ by reflectivity measurements.
In order to determine whether this 95 cm$^{-1}$ phonon 
is the driving force behind the dynamic DM interaction in this compound
normal mode calculations have to be performed.
To our knowledge this has not been done yet. 
Unfortunately, the limited $B_0$ range does not allow us to determine the dynamic DM-active $a$-axis optical phonon of T$_2$ triplet unambiguously.

\section{Conclusions}
We have reported the magnetic field and temperature dependence of two triplet modes in the chains of Sr$_{14}$Cu$_{24}$O$_{41}$
 with zero field excitation energies $\Delta_{1}=77.8$\,cm$^{-1}$ and $\Delta_{2}=87.7$\,cm$^{-1}$ at 4\,K. 
The triplet excitation at 77.8\,cm$^{-1}$ has not been observed before at the center of Brillouin zone.
Hence the presently accepted spin lattice model for the chains in Sr$_{14}$Cu$_{24}$O$_{41}$ is incomplete and must include a triplet excitation at 77.8 cm$^{-1}$ at $\mathbf{k}= 0$.
We propose a back-folding of the triplet branches (observed in INS) due to the doubling of the magnetic supercell from 5 to 10 chain units.
The determination of $g$ factors for free spins and triplets gave additional evidence
 that unpaired spins are in the chains and therefore the 5 chain unit periodicity cannot be retained.
Optical selection rules for the observed singlet-triplet transitions are consistent with the dynamic Dzyaloshinskii-Moriya transition mechanism.

\section{Acknowledgments}
Support by Estonian Science Foundation grants 4926, 5553 and 6138 is acknowledged. 
National High Magnetic Field Laboratory is supported by NSF Cooperative Agreement No. DMR-0084173, by the State of Florida, and by the DOE.
D.H.'s visit to the NHMFL was supported by travel scholarship V.05-06/13 from the Archimedes Foundation. 

\bibliographystyle{apsrev}

\begin{thebibliography}{27}
\expandafter\ifx\csname natexlab\endcsname\relax\def\natexlab#1{#1}\fi
\expandafter\ifx\csname bibnamefont\endcsname\relax
  \def\bibnamefont#1{#1}\fi
\expandafter\ifx\csname bibfnamefont\endcsname\relax
  \def\bibfnamefont#1{#1}\fi
\expandafter\ifx\csname citenamefont\endcsname\relax
  \def\citenamefont#1{#1}\fi
\expandafter\ifx\csname url\endcsname\relax
  \def\url#1{\texttt{#1}}\fi
\expandafter\ifx\csname urlprefix\endcsname\relax\def\urlprefix{URL }\fi
\providecommand{\bibinfo}[2]{#2}
\providecommand{\eprint}[2][]{\url{#2}}

\bibitem[{\citenamefont{McCarron et~al.}(1988)\citenamefont{McCarron,
  Subramanian, Calabrese, and Harlow}}]{mccarron88}
\bibinfo{author}{\bibfnamefont{E.~M.} \bibnamefont{McCarron}},
  \bibinfo{author}{\bibfnamefont{M.~A.} \bibnamefont{Subramanian}},
  \bibinfo{author}{\bibfnamefont{J.~C.} \bibnamefont{Calabrese}},
  \bibnamefont{and} \bibinfo{author}{\bibfnamefont{R.~L.}
  \bibnamefont{Harlow}}, \bibinfo{journal}{Materials Research Bulletin}
  \textbf{\bibinfo{volume}{23}}, \bibinfo{pages}{1355} (\bibinfo{year}{1988}),
  \urlprefix\url{http://www.sciencedirect.com/science/article/B6TXC-48JM1X4-6W%
/2/a2037421c176364e88aa5b51f123a03a}.

\bibitem[{\citenamefont{Siegrist et~al.}(1988)\citenamefont{Siegrist,
  Schneemeyer, Sunshine, Waszczak, and Roth}}]{siegrist88}
\bibinfo{author}{\bibfnamefont{T.}~\bibnamefont{Siegrist}},
  \bibinfo{author}{\bibfnamefont{L.~F.} \bibnamefont{Schneemeyer}},
  \bibinfo{author}{\bibfnamefont{S.~A.} \bibnamefont{Sunshine}},
  \bibinfo{author}{\bibfnamefont{J.~V.} \bibnamefont{Waszczak}},
  \bibnamefont{and} \bibinfo{author}{\bibfnamefont{S.}~\bibnamefont{Roth}},
  \bibinfo{journal}{Materials Research Bulletin} \textbf{\bibinfo{volume}{23}},
  \bibinfo{pages}{1429} (\bibinfo{year}{1988}),
  \urlprefix\url{http://www.sciencedirect.com/science/article/B6TXC-48DYH0H-5C%
/2/c159c7db0d435445d733466292452755}.

\bibitem[{\citenamefont{Uehara et~al.}(1996)\citenamefont{Uehara, Nagata,
  Akimitsu, Takahashi, Mori, and Kinoshita}}]{uehara96}
\bibinfo{author}{\bibfnamefont{M.}~\bibnamefont{Uehara}},
  \bibinfo{author}{\bibfnamefont{T.}~\bibnamefont{Nagata}},
  \bibinfo{author}{\bibfnamefont{J.}~\bibnamefont{Akimitsu}},
  \bibinfo{author}{\bibfnamefont{H.}~\bibnamefont{Takahashi}},
  \bibinfo{author}{\bibfnamefont{N.}~\bibnamefont{Mori}}, \bibnamefont{and}
  \bibinfo{author}{\bibfnamefont{K.}~\bibnamefont{Kinoshita}},
  \bibinfo{journal}{Journal of the Physical Society of Japan}
  \textbf{\bibinfo{volume}{65}}, \bibinfo{pages}{2764} (\bibinfo{year}{1996}),
  \urlprefix\url{http://dx.doi.org/10.1143/JPSJ.65.2764}.

\bibitem[{\citenamefont{Vuletic et~al.}(2006)\citenamefont{Vuletic,
  Korin-Hamzic, Ivek, Tomic, andM. Dressel, and Akimitsu}}]{vuletic06}
\bibinfo{author}{\bibfnamefont{T.}~\bibnamefont{Vuletic}},
  \bibinfo{author}{\bibfnamefont{B.}~\bibnamefont{Korin-Hamzic}},
  \bibinfo{author}{\bibfnamefont{T.}~\bibnamefont{Ivek}},
  \bibinfo{author}{\bibfnamefont{S.}~\bibnamefont{Tomic}},
  \bibinfo{author}{\bibfnamefont{B.~P.~G.} \bibnamefont{andM. Dressel}},
  \bibnamefont{and} \bibinfo{author}{\bibfnamefont{J.}~\bibnamefont{Akimitsu}},
  \bibinfo{journal}{Physics Reports} \textbf{\bibinfo{volume}{428}},
  \bibinfo{pages}{169} (\bibinfo{year}{2006}).

\bibitem[{\citenamefont{Zhang and Rice}(1988)}]{zhang88}
\bibinfo{author}{\bibfnamefont{F.~C.} \bibnamefont{Zhang}} \bibnamefont{and}
  \bibinfo{author}{\bibfnamefont{T.~M.} \bibnamefont{Rice}},
  \bibinfo{journal}{Phys. Rev. B} \textbf{\bibinfo{volume}{37}},
  \bibinfo{pages}{3759} (\bibinfo{year}{1988}).

\bibitem[{\citenamefont{Matsuda and Katsumata}(1996)}]{matsuda96}
\bibinfo{author}{\bibfnamefont{M.}~\bibnamefont{Matsuda}} \bibnamefont{and}
  \bibinfo{author}{\bibfnamefont{K.}~\bibnamefont{Katsumata}},
  \bibinfo{journal}{Phys. Rev. B} \textbf{\bibinfo{volume}{53}},
  \bibinfo{pages}{12201} (\bibinfo{year}{1996}).

\bibitem[{\citenamefont{Carter et~al.}(1996)\citenamefont{Carter, Batlogg,
  Cava, Krajewski, Peck, and Rice}}]{carter96}
\bibinfo{author}{\bibfnamefont{S.~A.} \bibnamefont{Carter}},
  \bibinfo{author}{\bibfnamefont{B.}~\bibnamefont{Batlogg}},
  \bibinfo{author}{\bibfnamefont{R.~J.} \bibnamefont{Cava}},
  \bibinfo{author}{\bibfnamefont{J.~J.} \bibnamefont{Krajewski}},
  \bibinfo{author}{\bibfnamefont{W.~F.} \bibnamefont{Peck},
  \bibfnamefont{Jr.}}, \bibnamefont{and} \bibinfo{author}{\bibfnamefont{T.~M.}
  \bibnamefont{Rice}}, \bibinfo{journal}{Phys. Rev. Lett.}
  \textbf{\bibinfo{volume}{77}}, \bibinfo{pages}{1378} (\bibinfo{year}{1996}).

\bibitem[{\citenamefont{Takigawa et~al.}(1998)\citenamefont{Takigawa, Motoyama,
  Eisaki, and Uchida}}]{takigawa98}
\bibinfo{author}{\bibfnamefont{M.}~\bibnamefont{Takigawa}},
  \bibinfo{author}{\bibfnamefont{N.}~\bibnamefont{Motoyama}},
  \bibinfo{author}{\bibfnamefont{H.}~\bibnamefont{Eisaki}}, \bibnamefont{and}
  \bibinfo{author}{\bibfnamefont{S.}~\bibnamefont{Uchida}},
  \bibinfo{journal}{Phys. Rev. B} \textbf{\bibinfo{volume}{57}},
  \bibinfo{pages}{1124} (\bibinfo{year}{1998}).

\bibitem[{\citenamefont{Matsuda et~al.}(1999)\citenamefont{Matsuda, Yosihama,
  Kakurai, and Shirane}}]{matsuda99}
\bibinfo{author}{\bibfnamefont{M.}~\bibnamefont{Matsuda}},
  \bibinfo{author}{\bibfnamefont{T.}~\bibnamefont{Yosihama}},
  \bibinfo{author}{\bibfnamefont{K.}~\bibnamefont{Kakurai}}, \bibnamefont{and}
  \bibinfo{author}{\bibfnamefont{G.}~\bibnamefont{Shirane}},
  \bibinfo{journal}{Phys. Rev. B} \textbf{\bibinfo{volume}{59}},
  \bibinfo{pages}{1060} (\bibinfo{year}{1999}).

\bibitem[{\citenamefont{Regnault et~al.}(1999)\citenamefont{Regnault, Boucher,
  Moudden, Lorenzo, Hiess, Ammerahl, Dhalenne, and Revcolevschi}}]{regnault99}
\bibinfo{author}{\bibfnamefont{L.~P.} \bibnamefont{Regnault}},
  \bibinfo{author}{\bibfnamefont{J.~P.} \bibnamefont{Boucher}},
  \bibinfo{author}{\bibfnamefont{H.}~\bibnamefont{Moudden}},
  \bibinfo{author}{\bibfnamefont{J.~E.} \bibnamefont{Lorenzo}},
  \bibinfo{author}{\bibfnamefont{A.}~\bibnamefont{Hiess}},
  \bibinfo{author}{\bibfnamefont{U.}~\bibnamefont{Ammerahl}},
  \bibinfo{author}{\bibfnamefont{G.}~\bibnamefont{Dhalenne}}, \bibnamefont{and}
  \bibinfo{author}{\bibfnamefont{A.}~\bibnamefont{Revcolevschi}},
  \bibinfo{journal}{Phys. Rev. B} \textbf{\bibinfo{volume}{59}},
  \bibinfo{pages}{1055} (\bibinfo{year}{1999}).

\bibitem[{\citenamefont{Eccleston et~al.}(1998)\citenamefont{Eccleston, Uehara,
  Akimitsu, Eisaki, Motoyama, and Uchida}}]{eccleston98}
\bibinfo{author}{\bibfnamefont{R.~S.} \bibnamefont{Eccleston}},
  \bibinfo{author}{\bibfnamefont{M.}~\bibnamefont{Uehara}},
  \bibinfo{author}{\bibfnamefont{J.}~\bibnamefont{Akimitsu}},
  \bibinfo{author}{\bibfnamefont{H.}~\bibnamefont{Eisaki}},
  \bibinfo{author}{\bibfnamefont{N.}~\bibnamefont{Motoyama}}, \bibnamefont{and}
  \bibinfo{author}{\bibfnamefont{S.-I.} \bibnamefont{Uchida}},
  \bibinfo{journal}{Phys. Rev. Lett.} \textbf{\bibinfo{volume}{81}},
  \bibinfo{pages}{1702} (\bibinfo{year}{1998}).

\bibitem[{\citenamefont{Fukuda et~al.}(2002)\citenamefont{Fukuda, Mizuki, and
  Matsuda}}]{fukuda02}
\bibinfo{author}{\bibfnamefont{T.}~\bibnamefont{Fukuda}},
  \bibinfo{author}{\bibfnamefont{J.}~\bibnamefont{Mizuki}}, \bibnamefont{and}
  \bibinfo{author}{\bibfnamefont{M.}~\bibnamefont{Matsuda}},
  \bibinfo{journal}{Phys. Rev. B} \textbf{\bibinfo{volume}{66}},
  \bibinfo{pages}{012104} (\bibinfo{year}{2002}).

\bibitem[{\citenamefont{van Smaalen}(2003)}]{smaalen03}
\bibinfo{author}{\bibfnamefont{S.}~\bibnamefont{van Smaalen}},
  \bibinfo{journal}{Physical Review B (Condensed Matter and Materials Physics)}
  \textbf{\bibinfo{volume}{67}}, \bibinfo{eid}{026101}
  (pages~\bibinfo{numpages}{1}) (\bibinfo{year}{2003}),
  \urlprefix\url{http://link.aps.org/abstract/PRB/v67/e026101}.

\bibitem[{\citenamefont{Braden et~al.}(2004)\citenamefont{Braden, Etrillard,
  Gukasov, Ammerahl, and Revcolevschi}}]{braden04}
\bibinfo{author}{\bibfnamefont{M.}~\bibnamefont{Braden}},
  \bibinfo{author}{\bibfnamefont{J.}~\bibnamefont{Etrillard}},
  \bibinfo{author}{\bibfnamefont{A.}~\bibnamefont{Gukasov}},
  \bibinfo{author}{\bibfnamefont{U.}~\bibnamefont{Ammerahl}}, \bibnamefont{and}
  \bibinfo{author}{\bibfnamefont{A.}~\bibnamefont{Revcolevschi}},
  \bibinfo{journal}{Physical Review B (Condensed Matter and Materials Physics)}
  \textbf{\bibinfo{volume}{69}}, \bibinfo{eid}{214426}
  (pages~\bibinfo{numpages}{6}) (\bibinfo{year}{2004}),
  \urlprefix\url{http://link.aps.org/abstract/PRB/v69/e214426}.

\bibitem[{\citenamefont{Etrillard et~al.}(2004)\citenamefont{Etrillard, Braden,
  Gukasov, Ammerahl, and Revcolevschi}}]{etrillard04}
\bibinfo{author}{\bibfnamefont{J.}~\bibnamefont{Etrillard}},
  \bibinfo{author}{\bibfnamefont{M.}~\bibnamefont{Braden}},
  \bibinfo{author}{\bibfnamefont{A.}~\bibnamefont{Gukasov}},
  \bibinfo{author}{\bibfnamefont{U.}~\bibnamefont{Ammerahl}}, \bibnamefont{and}
  \bibinfo{author}{\bibfnamefont{A.}~\bibnamefont{Revcolevschi}},
  \bibinfo{journal}{Physica C} \textbf{\bibinfo{volume}{403}},
  \bibinfo{pages}{290} (\bibinfo{year}{2004}).

\bibitem[{\citenamefont{Gotoh et~al.}(2003)\citenamefont{Gotoh, Yamaguchi,
  Takahashi, Akimoto, Goto, Onoda, Fujino, Nagata, and Akimitsu}}]{gotoh03}
\bibinfo{author}{\bibfnamefont{Y.}~\bibnamefont{Gotoh}},
  \bibinfo{author}{\bibfnamefont{I.}~\bibnamefont{Yamaguchi}},
  \bibinfo{author}{\bibfnamefont{Y.}~\bibnamefont{Takahashi}},
  \bibinfo{author}{\bibfnamefont{J.}~\bibnamefont{Akimoto}},
  \bibinfo{author}{\bibfnamefont{M.}~\bibnamefont{Goto}},
  \bibinfo{author}{\bibfnamefont{M.}~\bibnamefont{Onoda}},
  \bibinfo{author}{\bibfnamefont{H.}~\bibnamefont{Fujino}},
  \bibinfo{author}{\bibfnamefont{T.}~\bibnamefont{Nagata}}, \bibnamefont{and}
  \bibinfo{author}{\bibfnamefont{J.}~\bibnamefont{Akimitsu}},
  \bibinfo{journal}{Physical Review B (Condensed Matter and Materials Physics)}
  \textbf{\bibinfo{volume}{68}}, \bibinfo{eid}{224108}
  (pages~\bibinfo{numpages}{15}) (\bibinfo{year}{2003}),
  \urlprefix\url{http://link.aps.org/abstract/PRB/v68/e224108}.

\bibitem[{\citenamefont{Gotoh et~al.}(2006)\citenamefont{Gotoh, Yamaguchi,
  Eisaki, Nagata, and Akimitsu}}]{gotoh06}
\bibinfo{author}{\bibfnamefont{Y.}~\bibnamefont{Gotoh}},
  \bibinfo{author}{\bibfnamefont{I.}~\bibnamefont{Yamaguchi}},
  \bibinfo{author}{\bibfnamefont{H.}~\bibnamefont{Eisaki}},
  \bibinfo{author}{\bibfnamefont{T.}~\bibnamefont{Nagata}}, \bibnamefont{and}
  \bibinfo{author}{\bibfnamefont{J.}~\bibnamefont{Akimitsu}},
  \bibinfo{journal}{Physica C: Superconductivity}
  \textbf{\bibinfo{volume}{445-448}}, \bibinfo{pages}{107}
  (\bibinfo{year}{2006}),
  \urlprefix\url{http://www.sciencedirect.com/science/article/B6TVJ-4JW7WX2-R/%
2/68fe8e55cc884354ee699e4966d87ad9}.

\bibitem[{\citenamefont{Isobe et~al.}(2000)\citenamefont{Isobe, Onoda, Ohta,
  Izumi, Kimoto, Takayama-Muromachi, Hewat, and Ohoyama}}]{isobe00}
\bibinfo{author}{\bibfnamefont{M.}~\bibnamefont{Isobe}},
  \bibinfo{author}{\bibfnamefont{M.}~\bibnamefont{Onoda}},
  \bibinfo{author}{\bibfnamefont{T.}~\bibnamefont{Ohta}},
  \bibinfo{author}{\bibfnamefont{F.}~\bibnamefont{Izumi}},
  \bibinfo{author}{\bibfnamefont{K.}~\bibnamefont{Kimoto}},
  \bibinfo{author}{\bibfnamefont{E.}~\bibnamefont{Takayama-Muromachi}},
  \bibinfo{author}{\bibfnamefont{A.~W.} \bibnamefont{Hewat}}, \bibnamefont{and}
  \bibinfo{author}{\bibfnamefont{K.}~\bibnamefont{Ohoyama}},
  \bibinfo{journal}{Phys. Rev. B} \textbf{\bibinfo{volume}{62}},
  \bibinfo{pages}{11667} (\bibinfo{year}{2000}).

\bibitem[{\citenamefont{Gelle and Lepetit}(2004)}]{gelle04}
\bibinfo{author}{\bibfnamefont{A.}~\bibnamefont{Gelle}} \bibnamefont{and}
  \bibinfo{author}{\bibfnamefont{M.-B.} \bibnamefont{Lepetit}},
  \bibinfo{journal}{Physical Review Letters} \textbf{\bibinfo{volume}{92}},
  \bibinfo{eid}{236402} (pages~\bibinfo{numpages}{4}) (\bibinfo{year}{2004}),
  \urlprefix\url{http://link.aps.org/abstract/PRL/v92/e236402}.

\bibitem[{\citenamefont{Gelle and Lepetit}(2005)}]{gelle05}
\bibinfo{author}{\bibfnamefont{A.}~\bibnamefont{Gelle}} \bibnamefont{and}
  \bibinfo{author}{\bibfnamefont{M.-B.} \bibnamefont{Lepetit}},
  \bibinfo{journal}{The European Physics Journal B}
  \textbf{\bibinfo{volume}{46}}, \bibinfo{pages}{489} (\bibinfo{year}{2005}).

\bibitem[{\citenamefont{N\"ucker et~al.}(2000)\citenamefont{N\"ucker, Merz,
  Kuntscher, Gerhold, Schuppler, Neudert, Golden, Fink, Schild, Stadler
  et~al.}}]{nucker00}
\bibinfo{author}{\bibfnamefont{N.}~\bibnamefont{N\"ucker}},
  \bibinfo{author}{\bibfnamefont{M.}~\bibnamefont{Merz}},
  \bibinfo{author}{\bibfnamefont{C.~A.} \bibnamefont{Kuntscher}},
  \bibinfo{author}{\bibfnamefont{S.}~\bibnamefont{Gerhold}},
  \bibinfo{author}{\bibfnamefont{S.}~\bibnamefont{Schuppler}},
  \bibinfo{author}{\bibfnamefont{R.}~\bibnamefont{Neudert}},
  \bibinfo{author}{\bibfnamefont{M.~S.} \bibnamefont{Golden}},
  \bibinfo{author}{\bibfnamefont{J.}~\bibnamefont{Fink}},
  \bibinfo{author}{\bibfnamefont{D.}~\bibnamefont{Schild}},
  \bibinfo{author}{\bibfnamefont{S.}~\bibnamefont{Stadler}},
  \bibnamefont{et~al.}, \bibinfo{journal}{Phys. Rev. B}
  \textbf{\bibinfo{volume}{62}}, \bibinfo{pages}{14384} (\bibinfo{year}{2000}).

\bibitem[{\citenamefont{Osafune et~al.}(1997)\citenamefont{Osafune, Motoyama,
  Eisaki, and Uchida}}]{osafune97}
\bibinfo{author}{\bibfnamefont{T.}~\bibnamefont{Osafune}},
  \bibinfo{author}{\bibfnamefont{N.}~\bibnamefont{Motoyama}},
  \bibinfo{author}{\bibfnamefont{H.}~\bibnamefont{Eisaki}}, \bibnamefont{and}
  \bibinfo{author}{\bibfnamefont{S.}~\bibnamefont{Uchida}},
  \bibinfo{journal}{Phys. Rev. Lett.} \textbf{\bibinfo{volume}{78}},
  \bibinfo{pages}{1980} (\bibinfo{year}{1997}).

\bibitem[{\citenamefont{Kataev et~al.}(2001)\citenamefont{Kataev, Choi,
  Gr\"uninger, Ammerahl, B\"uchner, Freimuth, and Revcolevschi}}]{kataev01}
\bibinfo{author}{\bibfnamefont{V.}~\bibnamefont{Kataev}},
  \bibinfo{author}{\bibfnamefont{K.-Y.} \bibnamefont{Choi}},
  \bibinfo{author}{\bibfnamefont{M.}~\bibnamefont{Gr\"uninger}},
  \bibinfo{author}{\bibfnamefont{U.}~\bibnamefont{Ammerahl}},
  \bibinfo{author}{\bibfnamefont{B.}~\bibnamefont{B\"uchner}},
  \bibinfo{author}{\bibfnamefont{A.}~\bibnamefont{Freimuth}}, \bibnamefont{and}
  \bibinfo{author}{\bibfnamefont{A.}~\bibnamefont{Revcolevschi}},
  \bibinfo{journal}{Phys. Rev. B} \textbf{\bibinfo{volume}{64}},
  \bibinfo{pages}{104422} (\bibinfo{year}{2001}).

\bibitem[{\citenamefont{Klingeler et~al.}(2006)\citenamefont{Klingeler,
  Buchner, Choi, Kataev, Ammerahl, Revcolevschi, and Schnack}}]{klingeler06}
\bibinfo{author}{\bibfnamefont{R.}~\bibnamefont{Klingeler}},
  \bibinfo{author}{\bibfnamefont{B.}~\bibnamefont{Buchner}},
  \bibinfo{author}{\bibfnamefont{K.-Y.} \bibnamefont{Choi}},
  \bibinfo{author}{\bibfnamefont{V.}~\bibnamefont{Kataev}},
  \bibinfo{author}{\bibfnamefont{U.}~\bibnamefont{Ammerahl}},
  \bibinfo{author}{\bibfnamefont{A.}~\bibnamefont{Revcolevschi}},
  \bibnamefont{and} \bibinfo{author}{\bibfnamefont{J.}~\bibnamefont{Schnack}},
  \bibinfo{journal}{Physical Review B (Condensed Matter and Materials Physics)}
  \textbf{\bibinfo{volume}{73}}, \bibinfo{eid}{014426}
  (pages~\bibinfo{numpages}{6}) (\bibinfo{year}{2006}),
  \urlprefix\url{http://link.aps.org/abstract/PRB/v73/e014426}.

\bibitem[{\citenamefont{Cepas and Ziman}(2004)}]{cepas04}
\bibinfo{author}{\bibfnamefont{O.}~\bibnamefont{Cepas}} \bibnamefont{and}
  \bibinfo{author}{\bibfnamefont{T.}~\bibnamefont{Ziman}},
  \bibinfo{journal}{Physical Review B (Condensed Matter and Materials Physics)}
  \textbf{\bibinfo{volume}{70}}, \bibinfo{eid}{024404}
  (pages~\bibinfo{numpages}{8}) (\bibinfo{year}{2004}),
  \urlprefix\url{http://link.aps.org/abstract/PRB/v70/e024404}.

\bibitem[{\citenamefont{R{\~o}{\~o}m
  et~al.}(2004{\natexlab{a}})\citenamefont{R{\~o}{\~o}m, H{\"u}vonen, Nagel,
  Hwang, Timusk, and Kageyama}}]{room2004borate}
\bibinfo{author}{\bibfnamefont{T.}~\bibnamefont{R{\~o}{\~o}m}},
  \bibinfo{author}{\bibfnamefont{D.}~\bibnamefont{H{\"u}vonen}},
  \bibinfo{author}{\bibfnamefont{U.}~\bibnamefont{Nagel}},
  \bibinfo{author}{\bibfnamefont{J.}~\bibnamefont{Hwang}},
  \bibinfo{author}{\bibfnamefont{T.}~\bibnamefont{Timusk}}, \bibnamefont{and}
  \bibinfo{author}{\bibfnamefont{H.}~\bibnamefont{Kageyama}},
  \bibinfo{journal}{Phys. Rev. B} \textbf{\bibinfo{volume}{70}},
  \bibinfo{eid}{144417} (pages~\bibinfo{numpages}{8})
  (\bibinfo{year}{2004}{\natexlab{a}}),
  \urlprefix\url{http://link.aps.org/abstract/PRB/v70/e144417}.

\bibitem[{\citenamefont{R{\~o}{\~o}m
  et~al.}(2004{\natexlab{b}})\citenamefont{R{\~o}{\~o}m, H{\"u}vonen, Nagel,
  Wang, and Kremer}}]{room2004NaVa}
\bibinfo{author}{\bibfnamefont{T.}~\bibnamefont{R{\~o}{\~o}m}},
  \bibinfo{author}{\bibfnamefont{D.}~\bibnamefont{H{\"u}vonen}},
  \bibinfo{author}{\bibfnamefont{U.}~\bibnamefont{Nagel}},
  \bibinfo{author}{\bibfnamefont{Y.-J.} \bibnamefont{Wang}}, \bibnamefont{and}
  \bibinfo{author}{\bibfnamefont{R.~K.} \bibnamefont{Kremer}},
  \bibinfo{journal}{Phys. Rev. B} \textbf{\bibinfo{volume}{69}},
  \bibinfo{eid}{144410} (pages~\bibinfo{numpages}{19})
  (\bibinfo{year}{2004}{\natexlab{b}}),
  \urlprefix\url{http://link.aps.org/abstract/PRB/v69/e144410}.

\end{thebibliography}

\protect\newpage

\end{document}